# Modeling joint eating-out destination choices incorporating group-level impedance: A case study of the Greater Tokyo Area.

Chenglin Han[1], Lichen Luo[1], Giancarlos Parady[1], Kiyoshi Takami[1], Makoto Chikaraishi[2], Noboru Harata[3]

*Abstract*

Individuals undertake both solo and joint activities as part of their overall activity-travel patterns. Compared to work and maintenance activities, social and leisure activities differ in that they exhibit high levels of temporal and spatial flexibility. In this study we used data from an ego-centric social networks survey in the Greater Tokyo Area and follow-up group activity survey to estimate a joint eating-out destination choice model explicitly incorporating group-level impedance. Consistent with the literature, travel time has a large impact on destination choice as measured by its elasticity; however, the elasticities of group-level maximum, average and median travel times are larger than individual-level travel times. Furthermore, we show that incorporating group-level impedance increases model performance up to 49% against the ego-level impedance model, a substantial increase that underscores the need to incorporate group-level characteristics in travel behavior models.

**1. Introduction**

Leisure can be defined as the time spent out of work or domestic tasks on activities such as recreation, cultural events, sports and socializing (Ettema and Schwanen 2012b), and is one of the fastest-growing segments of trips in the 21st century (Caille et al. 2000). Given an increasing demand for more diversified functions of urban transportation system and a concern about this growth, mostly formulated as a reduction in miles travelled (Axhausen 2003), the study of leisure activities and related travel patterns is getting more attention. Furthermore, since the vast majority of leisure activities are undertaken in company of others, their dependence on the attributes, preferences and spatio-temporal constraints of others is non-ignorable.

   Although the activity-based approach was introduced around fifty years ago in the transportation field, research on leisure activity, especially leisure destination choice explicitly incorporation social network characteristics is still very limited. Against this background, the aim of this paper is to estimate an empirical destination choice model of joint-activities explicitly incorporating group-level impedance. Due to high diversity and variability of leisure activities, we narrowed the target to joint eating-out activities, one of the most common leisure activities with a highly social nature. We expect that explicitly considering the link between social networks and travel behavior could provide a better understanding of group behavior and consequently result in a model that outperforms traditional models which focus only on ego-level attributes and disregard group interactions.

---

[1] The University of Tokyo
[2] Hiroshima University
[3] Chuo University

The rest of this article is organized as follows: Chapter 2 provides an analysis of the current body of literature related to leisure activities. Chapter 3 describes the survey and data characteristics. Chapter 4 describes the methodology and modelling process. Chapter 5 discusses the estimation results and effect magnitude and Chapter 6 summarizes the article conclusions and proposed directions for further research.

## 2. Literature review

The inherently relational and social nature of leisure activities underscores the need to incorporate social networks in to better understand the relationship between activities and travel behavior. As a result, a growing number of studies are focusing on the connection between social networks and travel behavior. The majority of these studies have addressed (1) social network characteristics and (2) factors influencing interaction frequency among network members. Specifically, the negative relationship between ego-alter distance and face-to-face interaction frequency has been consistently reported (van den Berg, Arentze, and Timmermans 2009; Carrasco and Cid-Aguayo 2012; Larsen, Axhausen, and Urry 2006; Mok, Wellman, and Carrasco 2010) as well as the complementary effect of different ICT modes like phone (van den Berg, Arentze, and Timmermans 2012; Frei and Ohnmacht 2016), e-mail (van den Berg et al. 2009, 2012) and short message service (SMS) (van den Berg et al. 2009, 2012; Frei and Ohnmacht 2016) on face-to-face contact are reported.

Specifically addressing leisure activities, some light has been shed on various dimensions such as activity frequency (Parady, Frei, et al. 2021; Parady et al. 2019; Schilich et al. 2004; Sharmeen and Ettema 2010; Tarigan and Kitamura 2009), activity duration and associated travel time (Cools, Moons, and Wets 2010; Ren and Kwan 2009). Other studies have focused on factors driving individuals' propensity in participating in leisure activities and spatial setting (Sener et al. 2008), vehicle ownership and sociodemographic factors have been found to influence leisure consumption (Farber and Páez 2009; Parady et al. 2019; Tarigan and Kitamura 2009). However, destination choice in leisure travel has attracted very little attention. Since the vast majority of leisure activities are undertaken in company of others (Sharmeen and Ettema 2010), their dependence on the attributes, preferences and spatio-temporal constraints of others is non-ignorable. In fact, leisure travel is mostly for social purposes – to meet friends, relatives and acquaintances and a small share of leisure is solitary. For instance, 50% of leisure trips were taken for social interaction in Germany (Schilich et al. 2004) and Mikami (2002) found that the Japanese spend 31.7 and 9.3 hours socializing with household members and friends respectively, accounting for 24.4% of all hours in a week.

Ettema and Schwanen (2012b) introduced a relational approach for the analysis of leisure activities and travel and emphasized the dependency of leisure on the participation of other people, the influence of the characteristics of activity location and social networks. Akar, Clifton, and Doherty (2011) analyzed location choice associated to discretionary activities (in-home vs. out-of-home) and found that travel attributes like travel time and travel cost, affect to a greater extent the location of active activities (activities requiring the participant to engage in an activity, physically or mentally, in a way that affects the outcome, like spectator sports, exercising) than passive activities (where the participant does not affect the outcome, like going to movies) or social activities (activities where socializing and interacting with others are the primary aims). Kusumastuti et al. (2010) found the importance of companionship and efficiency in shopping which is associated with type of store, accessibility of car and bus is an important factor of location choices regarding shopping for fun. However, the number of studies on factors influencing location choice of leisure trips are still limited, in particular those focusing on joint leisure activities. While some studies have indeed focused on modeling joint activities, these studies rely on agent-based simulations (Arentze and Timmermans, 2008; Ronald, Arentze and Timmermans, 2012) and still require empirical data for parameter estimation and model validation. To the best of our knowledge no study has so far empirically evaluated destination choice of joint leisure activities explicitly incorporating group-level attributes, such as group-level impedance, a gap this study aims to fill.

## 3. Survey characteristics

The data used in this study comes from two surveys. The first is a probability survey of egocentric social networks in the Greater Tokyo Area. The second survey is a follow-up survey of the first one and focusing on group activities.

Regarding the first survey, its design and characteristics are reported in Parady, Takami & Harata (2021) but we will present a brief summary here for completeness. The survey objective was to measure social network characteristics and social interaction patterns in the Greater Tokyo Area (the Greater Tokyo Area is defined as the combination of the Tokyo Metropolis and the prefectures of Chiba, Saitama and Kanagawa.) In terms of sampling, generalized accessibility as defined by Parady, Loder & Axhausen (2017) was applied to stratify the Greater Tokyo Area into sextiles and one municipality per strata was sampled (Shinagawa ward, Matsudo city, Hino city, Kuki city, Odawara city and Kimitsu city). The total invited sample size was 1,000 and the number of samples per municipality was determined by the population share of each accessibility sextile. Followed the standards set by the American Association of Public Opinion Research (AAPOR, 2011), Response rate RR1 (returned, with complete responses) and RR2 (returned, with complete and partial responses) were 18.1% and 21.7%, respectively. The survey collected data on (i) respondents' (hereinafter ego) sociodemographic characteristics, and (ii) their egocentric networks measured via a name-generator and (iii) a sociogram to identify cliques within the network. To encourage respondents to complete the questionnaire, a digital JPY1,000 Amazon.co.jp gift card was offered to online respondents by e-mail while a physical gift card of equivalent value was mailed to respondents who completed the paper-based version. The reader is referred to Parady, Takami & Harata (2021) for specific details.

Regarding the second survey, out of original 217 respondents, respondents who agreed to participate in a follow-up survey (95 persons) were invited to respond to a survey aiming to capture attributes of joint eating-out activities conducted in early 2020[2]. Response rate RR1 and RR2 were 60% and 75.6% respectively. The same economic incentive was provided as in the first

---

[2] Although COVID-19 began to spread in Spring 2020, the survey submission deadline was March 31st, 2020 and 66% of the samples were already collected by March 26th, 2020 when the Tokyo Metropolis and the four adjacent prefectures in the capital region issued a voluntary stay-at-home request. Furthermore, the survey asked the respondents for last joint eating-out activity retrospectively. Thus, it is unlikely that COVID-19 affected the results.

survey. This resulted in an effective sample of 57 egos and 101 cliques.

The survey collected data on eating-out activities with each clique recorded in the original survey. The first section focused on the joint eating-out frequency and on the time interval since their last eating-out activity. The second section focused on eliciting specific information about last eating-out activity, including time, destination location, party members, other restaurants often visited together etc. The last section asked respondent to score the importance they place on different factors when choosing the destination for eating-out activity. Variable definitions are summarized in Table 1.

**Table 1.** Description of survey variables

| Variable name | Definition |
|---|---|
| **Clique information (collected for each clique)** | |
| Clique name | Type of clique, like family, colleague |
| Eating-out frequency | Frequency of eating-out with this clique |
| **Eating-out event information (collected for each clique)** | |
| Time since last eating-out | Time since last eating-out event with this clique |
| Day of week | If this event is held on weekend or at weekdays |
| Time of day | If this event is held in the morning, at noon or in the evening |
| Activity destination location | Address of event destination |
| Origin location (Ego only) | Longitude and latitude of origin for this event |
| Alternative activity locations | Location of other restaurants often visited with this clique (Maximum five locations) |
| Representative mode (Ego only) | Representative mode from origin to location |
| Other family member | If a family member not listed joined this event |
| Other friends | If a friend not listed joined this event |
| Number of alters | Number of alters within this clique joint this event |
| **Attitudes related to eating-out location choice** | |
| *Factors about alternate places (Five-point ordinal scale)* | |
| Visit frequency of alternate place | Visit frequency of places listed above |
| Convenience of alternate place | Feeling of convenience of places listed above |
| Rating of town | Personal preference of towns where places listed are in |
| *Other factors considered (seven points Likert scale)* | |
| Car access (myself) | A place where is accessible by car for me. |
| Transit access (myself) | A place where is accessible by transit for me. |
| NMM access (myself) | A place where is accessible by bicycle or foot for me. |
| Public reviews | Public review of a place |
| Car access (group) | A place where is accessible by car for group member |
| Transit access (group) | A place where is accessible by transit for group member |
| Non-motorized mode access (group) | A place where is accessible by bicycle or foot for group |
| Place rating (group) | Group member's preference of a place |
| Place atmosphere | Atmosphere of a place |
| Food quality | Food quality of a place |
| Place familiarity | If this place is familiar to me |
| Place I've never been to | If I've never been to this place |
| Town atmosphere | Atmosphere of town where this place is in |
| Town I've never been to | If I've never been to this town |
| Lots of shops in town | There are many shops in this town |
| Town rating (group) | Group member's preference of a town |
| Ease to go to After-party | Ease to go to attend other places after the first activity |
| Town isn't crowded | Town is not crowded |

## 3. Data characteristics

For conciseness, socio-demographic characteristics of the sample are summarized in Appendix A. Network size characteristics are summarized in Table 2. The average network size was 10.75 which is consistent with network sizes reported in previous Japanese studies (Parady, Takami and Harata, 2021).

**Table 2** Network size of egos (n=57)

| | Sample mean | Sample SD | Min | Max |
|---|---|---|---|---|
| Network size | | | | |
| All | 10.75 | 9.28 | 1 | 52 |
| Male | 8.48 | 6.40 | 1 | 26 |
| Female | 13.46 | 11.55 | 3 | 52 |
| 20-29 | 12.64 | 6.90 | 3 | 24 |
| 30-39 | 13.83 | 10.44 | 3 | 31 |
| 40-49 | 12.36 | 14.98 | 1 | 52 |
| 50-59 | 8.69 | 8.18 | 1 | 26 |
| Over 60 | 8.88 | 6.35 | 1 | 24 |

Ego-alter relational characteristics of returned sample are summarized in Table 3. Out of the 1448 alters with complete information in the original survey, data from 546 alters related to the 57 valid egos observations were used in the descriptive analysis. The geographical location of egos and alters is plotted in Figure 1. It can be seen that the density distribution of alters is to a certain consistent with the population density distribution of the Greater Tokyo Area.

**Table 3** Relational characteristics of egos and alters (n=546)

| Variable | Share (%) | | |
|---|---|---|---|
| | All | Ego is male | Ego is female |
| Homophily | | | |
| Gender | 70.64 | 67.68 | 72.13 |
| Occupation type | 49.58 | 50.43 | 48.87 |
| Age cohort | 37.98 | 35.89 | 38.75 |
| Civil status | 49.65 | 51.23 | 48.04 |
| Relationship length | | | |
| 1 year or less | 3.57 | 1.34 | 6.22 |
| 1-5 years | 15.62 | 16.24 | 14.88 |
| 5-10 years | 17.77 | 18.63 | 16.75 |
| Over 10 years | 63.04 | 63.79 | 62.15 |
| Relationship type | | | |
|   Immediate family | 23.23 | 27.15 | 18.56 |
|   Extended family | 10.46 | 6.84 | 14.78 |
|   Work | 15.21 | 17.47 | 12.51 |
|   School | 14.63 | 10.32 | 19.76 |
|   Neighbor | 5.59 | 6.88 | 4.06 |
|   Club or circle friend | 12.48 | 9.55 | 15.98 |
|   Other | 18.40 | 21.79 | 14.35 |
| Tie strength | | | |
|   Tier 1 (would discuss important problems with) | 32.50 | 31.55 | 33.64 |
|   Tier 2 (would ask for help in an emergency) | 27.71 | 30.17 | 24.78 |
|   Tier 1 and 2 | 37.79 | 37.28 | 38.47 |

| | Mean | Median | SD | Min | Max |
|---|---|---|---|---|---|
| Geodesic distance (km) | | | | | |
|   All | 170.79 | 12.14 | 996.77 | 0 | 10876 |
|   Male | 71.31 | 13.24 | 338.46 | 0 | 4604 |
|   Female | 237.42 | 12.07 | 1254.32 | 0 | 10876 |
| Contact frequency (1/year) | | | | | |
|   Face to face | 75.28 | 21.00 | 126.85 | 0 | 365 |
|   Phone | 27.32 | 2.00 | 75.08 | 0 | 365 |
|   Email/SMS | 32.45 | 0.00 | 86.12 | 0 | 365 |
|   SNS | 44.64 | 0.00 | 100.37 | 0 | 365 |

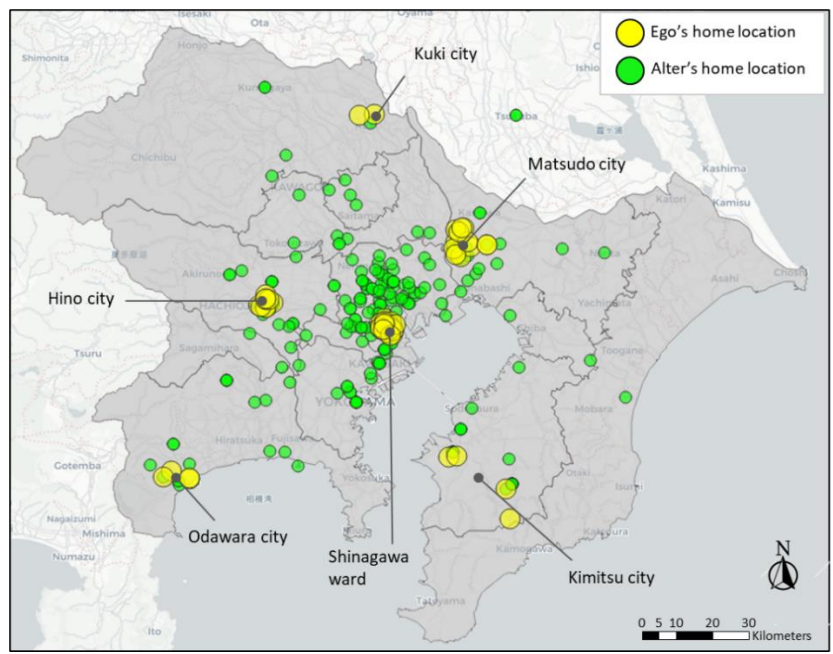

**Fig. 1** Spatial distribution of egos and alters in the sample.

**Characteristics of last joint activity by clique**

Characteristics of last eating-out activity for the 101 cliques reported in this survey are summarized in Tables 4 and 5, and the main findings from data distributions are as below:
- ➢ 54.5% of them were conducted at weekends and 45.5% of them were conducted on weekdays
- ➢ 42.6% and 30.7% of all reported activities happened at noon and in the evening
- ➢ One third of all cliques eat-out together 1-3 times/year and 4-11 times/year, respectively. Only a few groups eat-out very often (more than 4-6 times per week) or rarely (less than once a year) suggesting a moderate level of activity participation in the sample
- ➢ 95% of all activities were held with no more than 4 people with two-person-activities accounting for the largest proportion (67%)

Regarding the spatial distribution of activities, joint activity locations were concentrated within a 10 km range from ego and alters home locations (78.4% of all samples), egos' working places (63.6% of all samples reporting working address) and egos' trip origin locations (86.4% of all samples).

**Table 4 Descriptive statistics of last joint activity** (n=101)

| Variable | Sample share (%) |
|---|---|
| **The time interval between the response day and last eating-out event** | |
| less than 1 week | 13.86 |
| 1 week | 7.92 |
| 2 weeks | 9.90 |
| 3 weeks | 5.94 |
| 1 month | 18.81 |
| 2 months | 15.84 |
| More than 3 months | 27.72 |
| **Day of week of last event** | |
| Weekday | 45.54 |
| Weekend | 54.46 |
| **Day of week of last event** | |
| Noon | 42.57 |
| Early evening | 26.73 |
| Night | 30.69 |
| **Clique eating-out frequency** | |
| less than yearly | 7.92 |
| 1-3 times/year | 32.67 |
| 4-11 times/year | 31.68 |
| 1-2 times/month | 16.83 |
| 3-4 times/month | 4.95 |
| 1-3 times/week | 5.94 |
| 4-6 times/week | 0.00 |
| every day | 0.00 |
| **Number of participants (including ego)** | |
| 2 | 67.32 |
| 3 | 18.81 |
| 4 | 10.89 |
| 5 | 3.96 |
| 6 | 0.99 |
| **Representative mode from origin to location (ego only)** | |
| Transit | 46.53 |
| Bus | 0.99 |
| Car | 22.77 |
| Bike | 3.96 |
| Walk | 25.74 |

**Table 5** Distance between key locations and joint activity location (km)

| | mean | SD | median | min | max |
|---|---|---|---|---|---|
| Distance between home and destination (Ego and alters) | 11.128 | 13.568 | 6.599 | 0.062 | 83.066 |
| Distance between work and destination (Ego) | 14.394 | 14.551 | 9.231 | 0.067 | 61.211 |
| Distance between origin and destination (Ego) | 8.581 | 12.900 | 3.861 | 0.000 | 80.777 |

**Eating-out location-related attitudes of ego**

Regarding factors considered in the destination choice process, transit accessibility and evaluation of shops on both individual and group level, especially food quality and atmosphere are recognized as the most important. On the other hand, car accessibility and a variety-seeking attitude were less important (see Fig 2.)

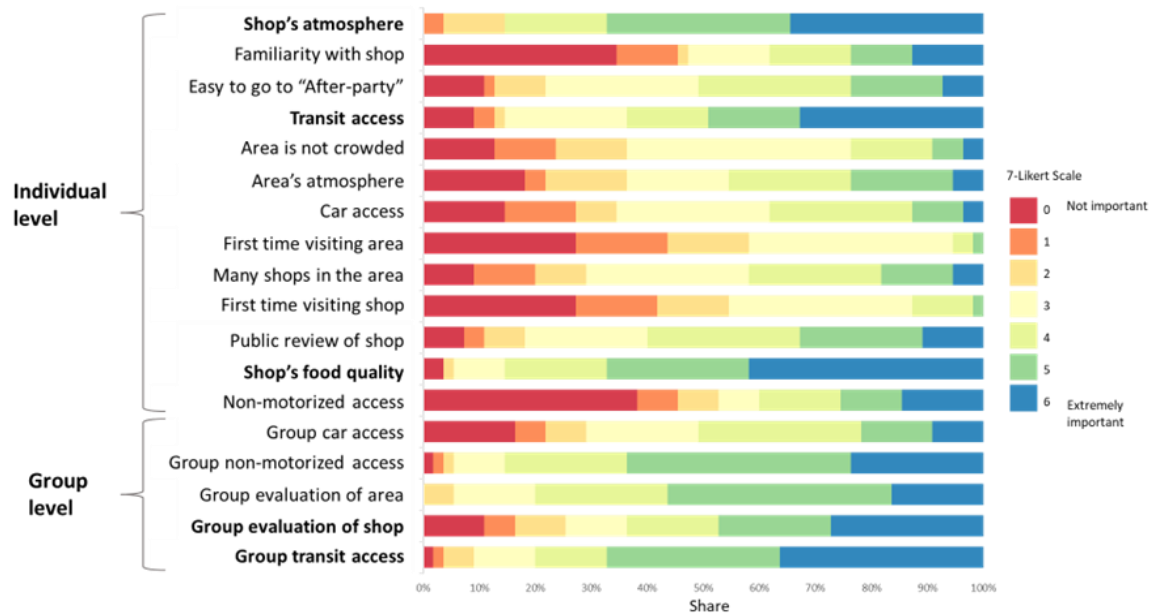

**Fig. 2** Rating score of importance of location attributes

## 4. Methodology

### 4.1. Data processing, model specification and estimation

To evaluate the factors affecting destination choice for joint eating-out activities and their effect magnitudes, a classical Multinomial Logit Model (MNL) is estimated. The spatial scale of analysis is the "small zones" (the smallest zone of the zone systems applied in Tokyo Metropolitan Region Person Trip Survey). Due to sample size limitations, we defined (i) the actual activity destination location and the (ii) alternative activity locations (see Table 1) as chosen alternatives, and use importance-based sampling to generate a feasible choice set in each choice situation (as explained below). This allows us to increase the number of choice situations from 101 to 261, and essentially means that we are estimating a model of potential destination choices within cliques. Regarding origin locations, we used the best information available in each case. When the dependent variable corresponds to the actual last activity conducted, for egos the origin corresponds to the actual origin location (which was observed in the data) while for alters it corresponds to their home locations. When the dependent variable corresponds to alternative locations, origins for egos and alters correspond to their home locations.

The 119 zones chosen in survey data are taken as the universal choice set. The average area of all alternative zones is 260 hectares and the standard deviation is 209.8 hectares. Model variable definitions are summarized in Table 6. Several model specifications were evaluated before narrowing down the explanatory variables to these ones. In contrast to other studies in the field which use broader range of activities when modeling destination choice (Scott & He, 2012), and to control for the high levels of variability of leisure activities, we explicitly narrowed down the target activity to eating-out in the survey design phase. This explains why adding additional land-use variables did not improve the model. Note that we have also tried to incorporate some other indicators of zonal attractiveness like public review scores of restaurants and have tried to estimate more complex size variables taking into by separating restaurants into several types like café, (Japanese) izakaya, bar etc., but these attempts did not improve the model.

**Table 6** Model variable description

| Variable | Description |
|---|---|
| Major station area | Dummy variable, takes value 1 if the zone is within a 1km radius from six major stations with a concentration of restaurants (Shinjuku, Shibuya, Ikebukuro, Ginza, Tokyo and Shimbashi). Data from the Economic Census for Business Activity of Japan in 2016 was used to get the geographical boundary information at the "small zone" level. 21 out of 119 alternative zones (17.64% alternatives) are within Major station areas. |
| Travel time | OD travel times were used as measures of impedance. Specifically, the following five different cost variables are empirically tested in the analysis:<br>Cost1: maximum travel time of group members<br>Cost2: minimum travel time of group members<br>Cost3: average travel time of group members<br>Cost4: median travel time of group members<br>Cost5: ego's travel time |
| Size variable | Logarithm of the number of restaurants in the zone. Alternative zones have on average 300 restaurants. Data was collected from a popular restaurant review site |

in Japan.

Regarding the calculation of travel times, since only ego travel mode choices were observed, we built a mode choice prediction model using ego data to predict alters' mode choices. Among several competing models, a simple logistic regression machine-learning classifier was chosen as it yielded the best performance (10-fold cross-validation percentage of correct predictions: 0.76).

For the sake of the stability of results, coefficients and standard errors are estimated via bootstrapping (r=50). In addition, for each iteration a different choice set was generated to account for the variability in estimates due to changes in the sampled alternatives. More specifically, the process is as follows:
1) Sample 261 observations with replacement from original dataset
2) For each observation, sample 20 zones from all 119 unique zones using importance-based sampling method where importance is defined using a gravity-type function of distance and attractiveness (number of restaurants). The sampling probability is calculated as follows (Ben-Akiva, 1981):

$$q_{in} = \widetilde{M}_i \cdot \exp\left(-\frac{2t_{in}}{\bar{t}}\right)$$

Where $\widetilde{M}_i$ is an approximate measure of size of destination zone $i$ (number of restaurants), $t_{in}$ is the estimated travel time between origin and destination $i$ for individual $n$, and $\bar{t}_{in}$ is the average travel time for all trips.
3) Estimate the MNL model.
4) Calculate direct elasticities for each variable and average over all zones.
5) Repeat 1) ~ 3) for $r$ times and take the average of values of direct elasticities

To evaluate the reproducibility of the model, defined as the ability of a model to maintain its predictive accuracy in a different sample from the same population (Parady, Ory, and Walker 2021), internal validation was conducted. Specifically, 10-fold cross-validation. Performance indicators were the percentage of correct predictions, and the fitting factor which is the average estimated choice probability of the chosen alternative (de Luca and Cantarella, 2009). Although the percentage of correct predictions is usually reported, the fitting factor is a better indicator, as it incorporates the probability with which the chosen alternative is predicted, as opposed to just considering the choice with the highest estimated probability irrespective of what this probability is.

## 5. Estimation results
The results of the estimation and validation of MNLs for five model specifications are presented in Table 7. Regarding model performance, internal validation results suggest that models explicitly considering group utility consistently outperform models that consider only ego's individual utility. Fitting factor improvements against the ego-only model ranged from 6.24% for the maximum distance model, to 49.33% for the average distance model, a considerable improvement in predictive ability.

In terms of estimated effects, destination choice is influenced by (1) the number of restaurants in the area as a measure of attractiveness and (2) travel distance of all members of the party. Direct elasticities of travel cost and number of restaurants are reported in Table 8. The elasticity of travel cost is largest for the model that uses average travel time among group members as an impedance measure, indicating an average 3.88% reduction in choice probability given a 1% increase in the average distance among group members, a very large effect.

Table 7 MNL model results for joint eating-out activity destination choice

|  | Maximum distance model | Minimum distance model | Average distance model | Median distance model | Individual distance model |
|---|---|---|---|---|---|
| **Estimated coefficients** | | | | | |
| Major station area | 0.0379 | 0.1206 | 0.0093 | 0.0488 | 0.0507 |
| Size variable: ln(*Number of Restaurants*) | **0.3225** | **0.6735** | **0.5590** | **0.4898** | **0.5137** |
| Cost1: maximum time of group | **-0.1262** | | | | |
| Cost2: minimum time of group | | **-0.3080** | | | |
| Cost3: average time of group | | | **-0.2943** | | |
| Cost4: median time of group | | | | **-0.2343** | |
| Cost5: ego's individual time | | | | | **-0.1517** |
| **Goodness of fit** | | | | | |
| Num. observations | 261 | 261 | 261 | 261 | 261 |
| Rho-squared | 0.1621 | 0.2346 | 0.2672 | 0.2430 | 0.1757 |
| Adjusted rho-squared | 0.1581 | 0.2304 | 0.2631 | 0.2389 | 0.1716 |
| **Validation performance (10-fold cross validation)** | | | | | |
| Percentage of correct prediction | 23.70 | 36.30 | 32.59 | 33.33 | 24.44 |
| Increase against individual model | -3.03 | 48.53 | 33.35 | 36.37 | - |
| Fitting factor | 16.67 | 20.77 | 23.43 | 21.57 | 15.69 |
| Increase against individual model | 6.24 | 32.38 | 49.33 | 37.48 | - |

*Coefficients statistically significant above the 0.10 level in bold*

Table 8. Average direct elasticity of significant variables of all alternative zones

|  | Maximum distance model | Minimum distance model | Average distance model | Median distance model | Individual distance model |
|---|---|---|---|---|---|
| ln(*Number of Res*) | 0.15 | 0.44 | 0.30 | 0.32 | 0.40 |
| 95% C.I. | 0.14 to 0.16 | 0.43 to 0.44 | 0.29 to 0.31 | 0.31 to 0.33 | 0.39 to 0.40 |
| Maximum distance of group | -3.41 | | | | |
| 95% C.I. | -3.44 to -3.38 | | | | |
| Minimum distance of group | | -1.97 | | | |
| 95% C.I. | | -1.98 to -1.96 | | | |
| Average distance of group | | | -3.88 | | |
| 95% C.I. | | | -3.91 to -3.85 | | |
| Median distance of group | | | | -3.24 | |
| 95% C.I. | | | | -3.26 to -3.22 | |
| Individual distance | | | | | -2.30 |
| 95% C.I. | | | | | -2.31 to -2.28 |

The second largest was the maximum travel time among group members. This suggests that groups consider locations that are on average convenient to all members and are more sensitive to the most inconvenienced member of the party and might weight his inconvenience when making a choice. Direct elasticities of number of restaurants indicate an average 0.15%-0.44% increase in choice probability given a 1% increase in the number of restaurants. Fig. 3 plots choice probability relative to travel time, and illustrates the greatest reduction in choice probability by the increase of average travel time of the whole party members.

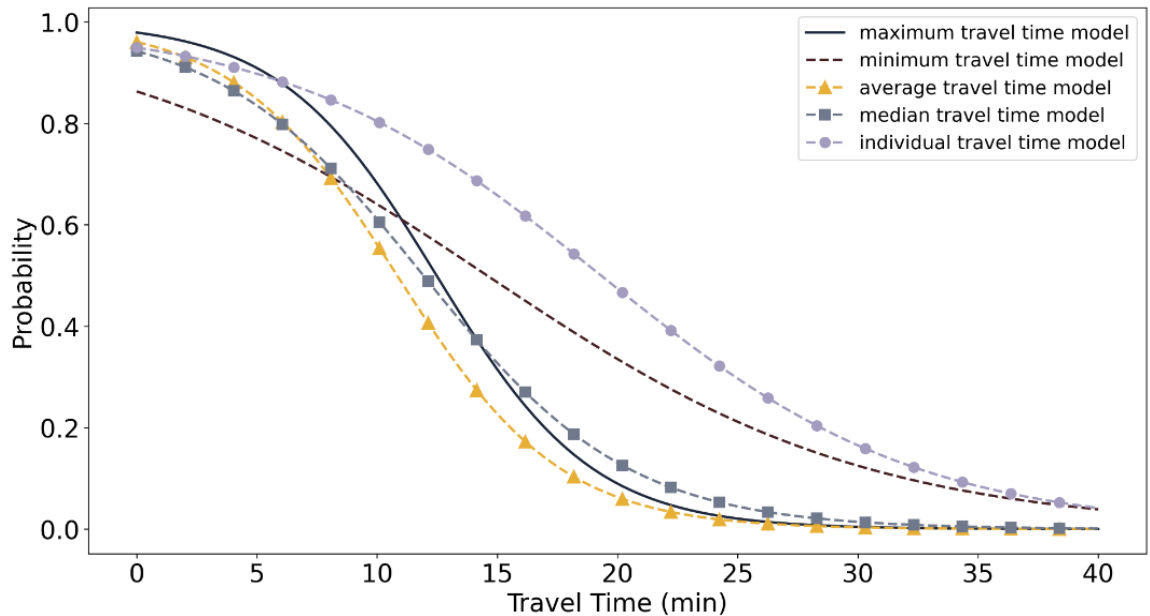

**Fig. 3** Plot of simulation on probability to travel time.
Number of restaurants was fixed to the average value of 300.

### 5.3. Effect magnitude in different market segments

To explore effect differences given group or event characteristics, we conducted a segmentation analysis. Note that due to sample limitations, segmentation was conducted the segmentation one category at a time. Six standards are applied for segmentation as shown in Table 9. Comparison of size effects on average travel time model are shown in Figures 4 to 8. Note that average participation rate of clique members in the last activity conducted was 94%, as such, the segmentation should be valid for both the "actual activity" and "frequently visited places with that clique".

Generally speaking, the effect size of travel time on choice probability differs in different segmentations. However, the effect difference of number of restaurants across different segments is essentially close to zero (see the differences in the y axis) so only the differences in travel time will be discussed.

Results suggest that mostly-elder groups are more sensitive to travel time (Fig. 4), which might be related to more limited levels of physical mobility compared to their younger counterparts. Groups with less than 5 members are also more sensitive to travel time (Fig. 6), indicating that for larger groups, group members might be more willing to consider the benefits of the group as a whole, even if it results in longer distances on average. No effect differences were observed given group gender composition or residential location (Fig. 5, Fig. 7).

Groups composed mostly of members that have known each other for more than 5 years are more sensitive to travel time (Fig. 8). It is possible that for groups that know each other longer, priority is given to some extent to companionship than the actual destination, hence putting more weight on proximity. Further studies are necessary, however, to validate these results and shed some more light on the mechanisms behind the observed effect differences.

Table 9. Type and specification of segmentation

| Segmentation label | Classification | |
|---|---|---|
| Age | Highest ratio of age group | Age<60 (n=165) |
| | | Age>=60 (n=96) |
| Gender | Highest ratio of gender group: | Female>0.5 (n=141) |
| | | Male>0.5 (n=120) |
| Party size | Party size | Size<=4 (n=153) |
| | | Size>4 (n=108) |
| Time of event | Time when the eating-out activity is held: | Noon (n=123) |
| | | Evening (n=138) |
| | Day when the eating-out activity is held | Weekend (n=159) |
| | | Weekday (n=102) |
| Ego's home is in 23 wards | Whether Ego's home location is in 23 wards | Yes (n=134) |
| | | No (n=127) |
| Length of relationship | Highest ratio of length of relationship between ego and alters: | <5 years (n=78) |
| | | >5 years (n=183) |

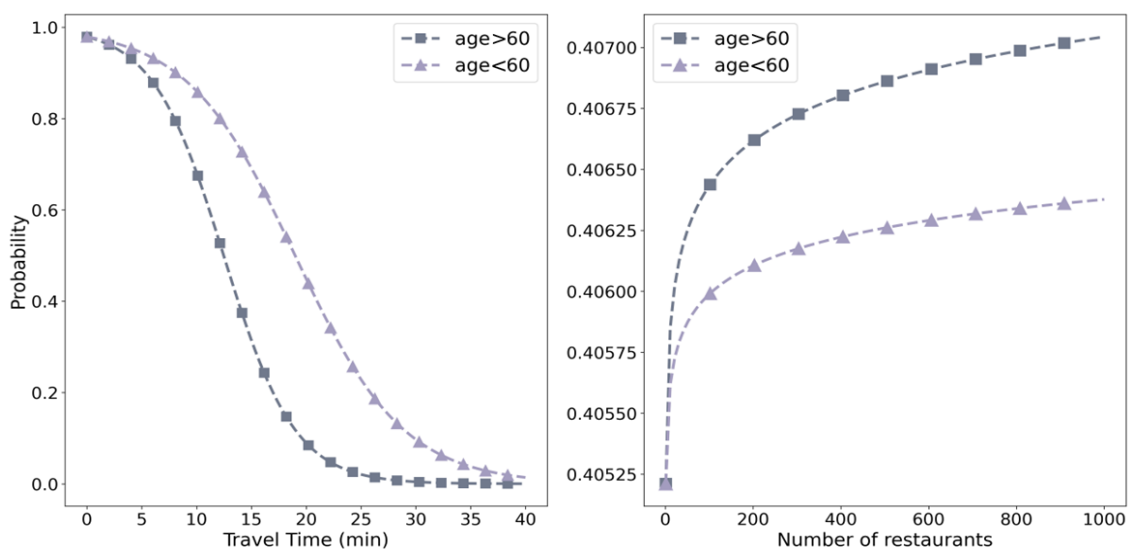

Fig. 4 Plot of simulation on probability to travel time and number of restaurants in different age groups.

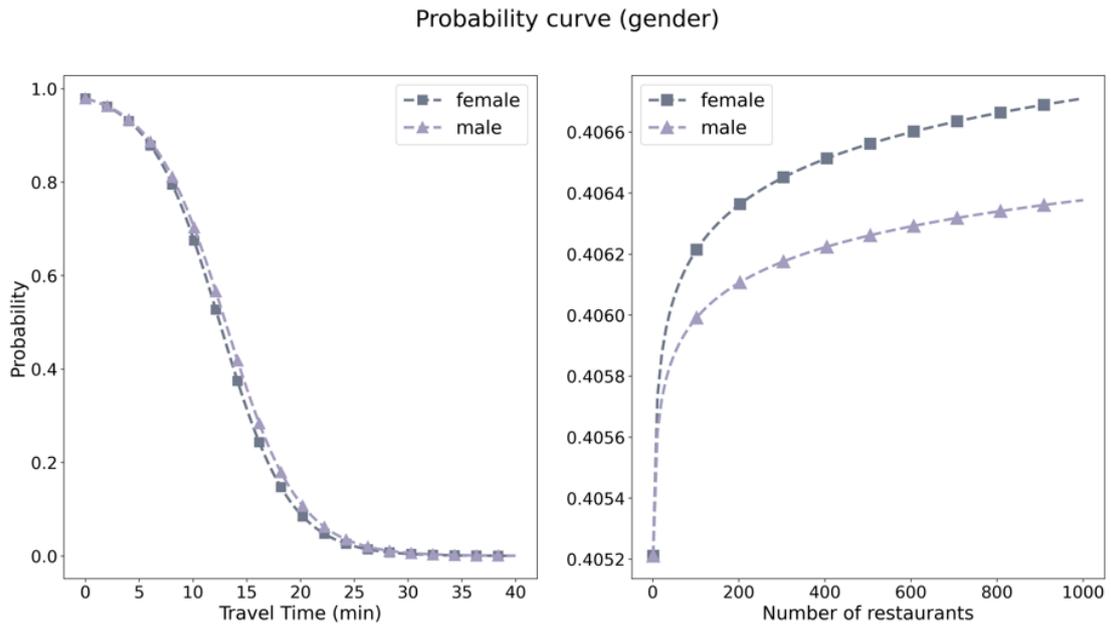

**Fig. 5** Plot of simulation on probability to travel time and number of restaurants in different gender groups

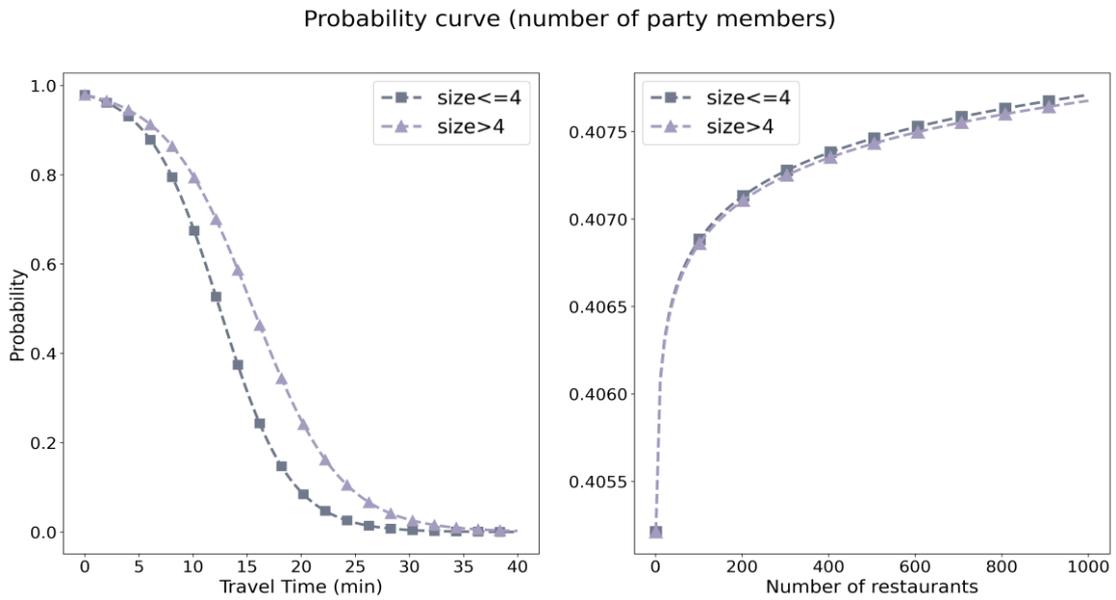

**Fig. 6** Plot of simulation on probability to travel time and number of restaurants in groups with different number of party members

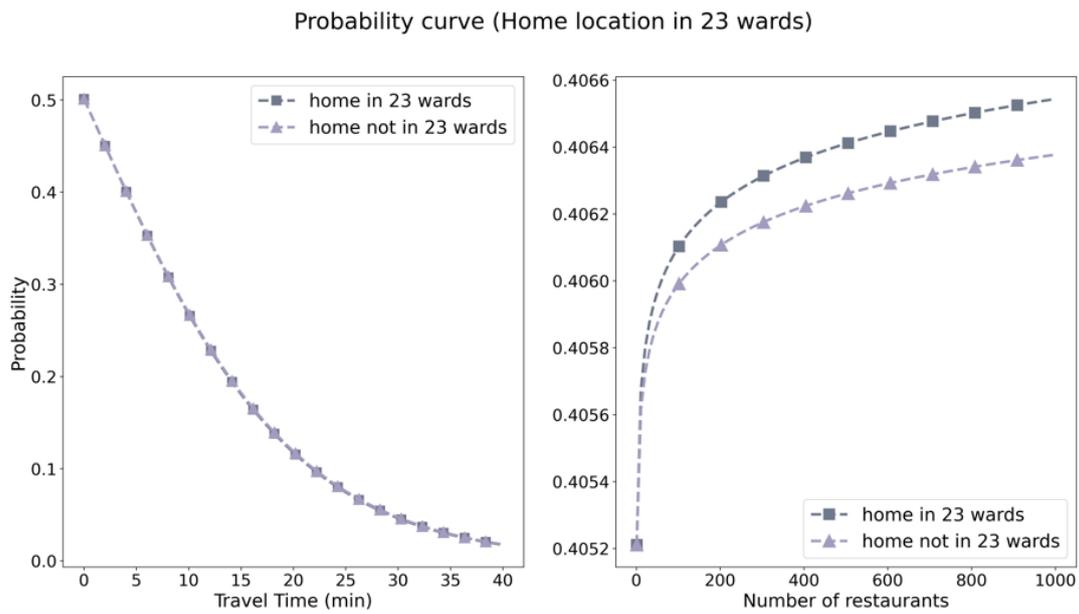

**Fig. 7** Plot of simulation on probability to travel time and number of restaurants in groups with ego living with Tokyo 23 wards or not

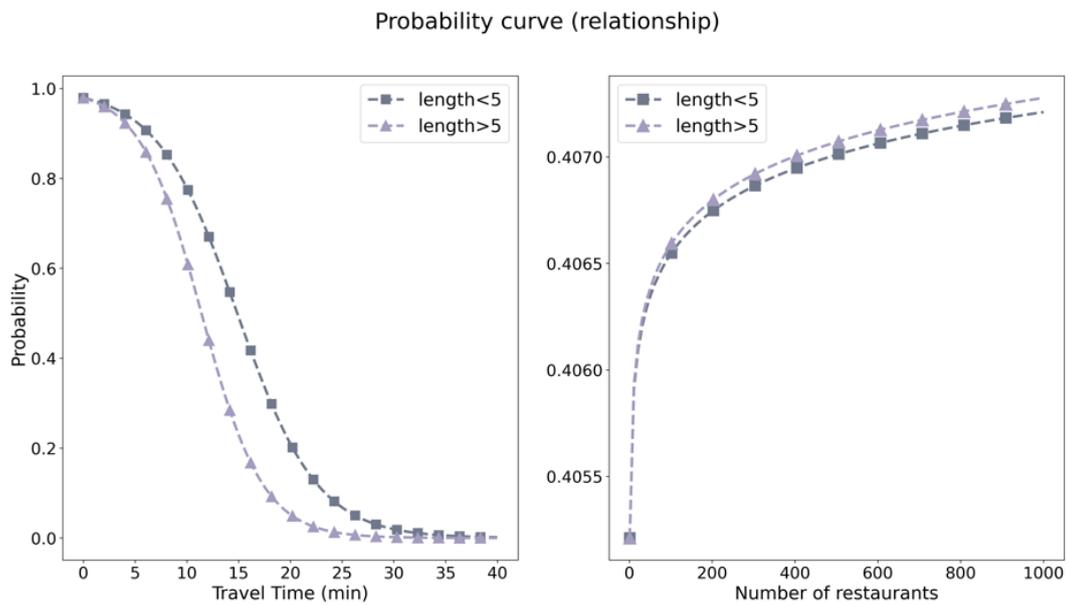

**Fig. 8** Plot of simulation on probability to travel time and number of restaurants in groups with party held within cliques of different relationship length

**6. Conclusions**

In this study we used data from an ego-centric social networks survey in the Greater Tokyo Area and follow-up group activity survey to estimate a joint eating-out destination choice model explicitly incorporating group-level impedance. Estimation results showed that models incorporating group utility markedly outperformed the individual utility model. Results also showed that the elasticities of group-level maximum, average and median travel times are larger than individual-level travel times, with the largest effect observed for the average-distance model. These findings underscore the need to incorporate social network characteristics in travel behavior models to improve the predictive ability of destination choice models.

In terms of the limitations of this study. First, due to high response burden associated with collecting information on both social networks and travel behavior at the clique level, the effective sample size is small. Survey methods aimed at reducing response burden are a welcomed contribution. However, it is important to highlight that in spite of this limitation, this is, to the best of our knowledge the first study that explicitly incorporates group-level impedance in destination choice models using empirical data. Second, information of social networks is based on ego self-reporting, so the information provided on group members is limited to what ego can recall, a key limitation of ego-centric studies. New survey methodologies are required to properly observe the decision-making process of groups as well as the spatio-temporal constraints, not only of ego, but of all group members. In this regard, the work of Parady, Oyama & Chikaraishi (2022) is a step forward in this direction.

**References**

Akar, Gulsah, Kelly J. Clifton, and Sean T. Doherty. 2011. "Discretionary Activity Location Choice: In-Home or out-of-Home?" *Transportation* 38(1). doi: 10.1007/s11116-010-9293-x.

Axhausen, K. W. 2003. "Social Networks and Travel : Some Hypotheses." *Arbeitsbericht Verkehrs Und Raumplanung* 197(December).

Arentze, T., & Timmermans, H. 2008. Social networks, social interactions, and activity-travel behavior: a framework for microsimulation. Environment and Planning B: Planning and Design, 35(6), 1012-1027.

Ben-Akiva, M., Watanatada, T., Manski, C., Mcfadden, D. 1981. Applications of a Continuous Spatial Choice Logit Model.

van den Berg, Pauline, Theo A. Arentze, and Harry J. P. Timmermans. 2009. "Size and Composition of Ego-Centered Social Networks and Their Effect on Geographic Distance and Contact Frequency." *Transportation Research Record* (2135). doi: 10.3141/2135-01.

van den Berg, Pauline, Theo Arentze, and Harry Timmermans. 2012. "A Multilevel Path Analysis of Contact Frequency between Social Network Members." *Journal of Geographical Systems* 14(2). doi: 10.1007/s10109-010-0138-0.

Caille, Pierre, Michel Kammermann, Ulrich Seewer, Kurt Infanger, Schweiz Bundesamt für Statistik, Schweiz Bundesamt für Raumentwicklung, des transports Suisse Département fédéral de l'environnement, Suisse Département fédéral de l'intérieur, Institut d'études de marché et de recherches sociales (Lausanne) LINK, Institut für Markt-und Sozialforschung (Luzern) LINK, Suisse Office fédéral de la statistique, and Suisse Office fédéral du développement territorial. 2000. *Mobilität in Der Schweiz Ergebnisse Des Mikrozensus 2000 Zum Verkehrsverhalten*.

Carrasco, Juan Antonio, and Beatriz Cid-Aguayo. 2012. "Network Capital, Social Networks, and Travel: An Empirical Illustration from Concepción, Chile." *Environment and Planning A* 44(5). doi: 10.1068/a43222.

Cools, Mario, Elke Moons, and Geert Wets. 2010. "Assessing the Impact of Public Holidays on Travel Time Expenditure: Differentiation by Trip Motive." *Transportation Research Record* (2157). doi: 10.3141/2157-04.

de Luca, S., Cantarella, G.E., 2009. Validation and comparison of choice models. In: Saleh, W.,

Sammer, G. (Eds.), Travel Demand Management and Road User Pricing: Success, Failure and Feasibility. Ashgate publications, pp. 37–58.

Ettema, Dick, and Tim Schwanen. 2012b. "A Relational Approach to Analysing Leisure Travel." *Journal of Transport Geography* 24:173–81. doi: 10.1016/j.jtrangeo.2012.01.023.

Farber, Steven, and Antonio Páez. 2009. "My Car, My Friends, and Me: A Preliminary Analysis of Automobility and Social Activity Participation." *Journal of Transport Geography* 17(3). doi: 10.1016/j.jtrangeo.2008.07.008.

Frei, Andreas, and Timo Ohnmacht. 2016. "Egocentric Networks in Zurich: Quantitative Survey Development, Data Collection and Analysis." in *Social Networks and Travel Behaviour*.

Kowald, Matthias, and Kay W. Axhausen. 2012. "Focusing on Connected Personal Leisure Networks: Selected Results from a Snowball Sample." *Environment and Planning A* 44(5). doi: 10.1068/a43458.

Kusumastuti, Diana, Els Hannes, Davy Janssens, Geert Wets, and Benedict G. C. Dellaert. 2010. "Scrutinizing Individuals' Leisure-Shopping Travel Decisions to Appraise Activity-Based Models of Travel Demand." *Transportation* 37(4). doi: 10.1007/s11116-010-9272-2.

Larsen, Jonas, Kay W. Axhausen, and John Urry. 2006. "Geographies of Social Networks: Meetings, Travel and Communications." *Mobilities* 1(2). doi: 10.1080/17450100600726654.

Mikami, S. 2002. "Internet Use and Sociability in Japan." *It & Society.* 242–50.

Mok, Diana, Barry Wellman, and Juan Carrasco. 2010. "Does Distance Matter in the Age of the Internet?" *Urban Studies* 47(13). doi: 10.1177/0042098010377363.

Parady, Giancarlos, Andreas Frei, Matthias Kowald, Sergio Guidon, Michael Wicki, Pauline van den Berg, Juan-Antonio Carrasco, Theo Arentze, Harry Timmermans, Barry Wellman, Kiyoshi Takami, Noboru Harata, and Kay Axhausen. 2021. "A Comparative Study of Social Interaction Frequencies among Social Network Members in Five Countries." *Journal of Transport Geography* 90:102934. doi: 10.1016/j.jtrangeo.2020.102934.

Parady, Giancarlos, David Ory, and Joan Walker. 2021. "The Overreliance on Statistical Goodness-of-Fit and under-Reliance on Model Validation in Discrete Choice Models: A Review of Validation Practices in the Transportation Academic Literature." *Journal of Choice Modelling* 38.

Parady, Giancarlos Troncoso, Genki Katayama, Hiromu Yamazaki, Tatsuki Yamanami, Kiyoshi Takami, and Noboru Harata. 2019. "Analysis of Social Networks, Social Interactions, and out-of-Home Leisure Activity Generation: Evidence from Japan." *Transportation* 46(3). doi: 10.1007/s11116-018-9873-8.

Parady, G. T., A. Loder, and K. W. Axhausen. 2017. "Heteroegenous Travel Activity Patterns in Japan: Accounting for Inter-Dependencies in Mobility Tool Use." Transportation Research Board Meeting.

Parady, Giancarlos, Kiyoshi Takami, and Noboru Harata. 2021. "Egocentric Social Networks and Social Interactions in the Greater Tokyo Area." Transportation 48(2):831–56. doi:10.1007/s11116-020-10079-y.

Parady, Giancarlos, Oyama Yuki, Chikaraishi Makoto. 2022. Understanding the joint decision-making process of leisure destination choices: Exploring new methodologies. Presented at the 16th International Conference on Travel Behavior Research, Santiago, Chile, December 11-15.

Ren, Fang, and Mei Po Kwan. 2009. "The Impact of the Internet on Human Activity-Travel Patterns: Analysis of Gender Differences Using Multi-Group Structural Equation Models." *Journal of Transport Geography* 17(6). doi: 10.1016/j.jtrangeo.2008.11.003.

Ronald, N., Arentze, T., & Timmermans, H. (2012). Modeling social interactions between individuals

for joint activity scheduling. Transportation research part B: methodological, 46(2), 276-290.

Saleh, Wafaa, and Gerd Sammer. 2009. "Travel Demand Management and Road User Pricing: Success, Failure and Feasibility." in *Travel Demand Management and Road User Pricing: Success, Failure and Feasibility*.

Schilich, ROBERT, STEFAN SCHÖNFELDER, SUSAN HANSON, and KAY W. AXHAUSEN. 2004. "Structures of Leisure Travel: Temporal and Spatial Variability." *Transport Reviews* 24(2):219–37. doi: 10.1080/0144164032000138742.

Scott, D. M., & He, S. Y. (2012). Modeling constrained destination choice for shopping: a GIS-based, time-geographic approach. Journal of Transport Geography, 23, 60-71.

Sener, Ipek N., Rachel B. Copperman, Ram M. Pendyala, and Chandra R. Bhat. 2008. "An Analysis of Children's Leisure Activity Engagement: Examining the Day of Week, Location, Physical Activity Level, and Fixity Dimensions." in *Transportation*. Vol. 35.

Sharmeen, Fariya, and Dick Ettema. 2010. "Whom to Hang out with and Where? Analysis of the Influence of Spatial Setting on the Choice of Activity Company." *12th WCTR Conference, 11–15 July, 2010, Lisbon, Portugal.* 18.

Stauffacher, Michael, Robert Schlich, K. W. Axhausen, and R. Scholz. 2005. "The Diversity of Travel Behaviour: Motives and Social Interactions in Leisure Time Activities." *Arbeitsberichte Verkehr-Und Raumplanung, 30x, IVT, ETH Zurich, Zurich* (November).

Tarigan, Ari K. M., and Ryuichi Kitamura. 2009. "Week-to-Week Leisure Trip Frequency and Its Variability." *Transportation Research Record: Journal of the Transportation Research Board* 2135(1):43–51. doi: 10.3141/2135-06.

The American Association for Public Opinion Research: Standard Definitions: Final Dispositions of Case Codes and Outcome Rates for Surveys, 9th edn. AAPOR, Lenexa (2011)

**Appendix A:** Socio-demographic characteristics of egos (n=57)

| Variable | Sample share (%) | Population share (%) |
|---|---|---|
| **Collection mode** | | |
| Web | 78.95 | |
| Paper | 21.05 | |
| **Gender** | | |
| Male | 54.39 | 49.24 |
| Female | 45.61 | 50.76 |
| **Age** | | |
| 20-29 | 19.30 | 13.28 |
| 30-39 | 10.53 | 16.56 |
| 40-49 | 19.30 | 19.55 |
| 50-59 | 22.81 | 14.62 |
| 60 and over | 28.07 | 35.99 |
| **Civil status** | | |
| Divorced | 7.02 | |
| In a relationship | 5.26 | |
| Married with children | 47.37 | |
| Married, no children | 8.77 | |
| Single | 28.07 | |
| Widowed | 1.75 | |
| Other | 1.75 | |
| **Employment** | | |
| Full time | 56.14 | |
| Home maker | 7.02 | |
| Freelance | 1.75 | |
| Part time | 12.28 | |
| Student | 8.77 | |
| Unemployed/retired | 8.77 | |
| Other | 5.26 | |
| **Education** | | |
| High-school and under | 24.56 | |
| Technical degree | 28.07 | |
| University degree and higher | 47.37 | |
| **Household income** | | |
| Not disclosed | 15.79 | 6.40 |
| Under 3 million | 15.79 | 21.50 |
| 3-10 million | 54.39 | 60.70 |
| Over 10 million | 14.04 | 11.40 |

| | Sample mean | Sample SD | Min | Max |
|---|---|---|---|---|
| Household size | 2.79 | 1.22 | 1 | 8 |

Population shares from 2015 Census.

Household income data was calculated using data from the 2013 Housing and Land Survey.